\documentclass[sigconf]{acmart}
\usepackage{xspace}

\AtBeginDocument{
  }

\setcopyright{acmlicensed}
\copyrightyear{2025}
\acmYear{2025}
\acmDOI{XXXXXXX.XXXXXXX}

\acmConference[EASE 2025]{The 29th International Conference on Evaluation and Assessment in Software Engineering}{17–20 June, 2025}{Istanbul, Türkiye}

\acmISBN{XXXXXXX.XXXXXXX}

\newcommand\numberToBeChecked[1]{\textcolor{black}{#1}}

\newcommand{\nbPapersFromAllEngines}{\numberToBeChecked{733 }\xspace}
\newcommand{\nbPapersFromAllEnginesWithoutDuplicates}{\numberToBeChecked{631}\xspace}

\newcommand{\nbPapersAfterInclusionCriteria}{\numberToBeChecked{42}\xspace}
\newcommand{\nbPapersFinalSelection}{\numberToBeChecked{42} \xspace}
\usepackage{xcolor}
\usepackage{tabularx}
\usepackage{tcolorbox}
\tcbuselibrary{skins}
\newenvironment{summarybox}
{\begin{tcolorbox}
[enhanced,arc=0mm,colback=gray!10,frame hidden,overlay unbroken={
    \draw[thick,black] (interior.north west)--(interior.south west);
},left=2pt,right=0pt,top=0pt,bottom=0pt,before={\noindent},after={}]}
{\end{tcolorbox}}
\newenvironment{summary}
{\noindent\begin{summarybox} Summary:}
{\end{summarybox}}

\begin{document}

\title{Understanding Underrepresented Groups in Open Source Software}

\author{Reydne Santos}
\orcid{https://orcid.org/0009-0006-3510-0521}
\affiliation{
  \institution{Universidade Federal de Pernambuco}
  \city{Recife}
  \state{Pernambuco}
  \country{Brazil}}
\email{rbs8@cin.ufpe.br}

\author{Rafa Prado}
\orcid{https://orcid.org/0000-0001-5851-3229}
\affiliation{
  \institution{Universidade Federal de Pernambuco}
  \city{Recife}
  \country{Brazil}}
\email{rps4@cin.ufpe.br}

\author{Ana Paula de Holanda Silva}
\orcid{https://orcid.org/0000-0003-1186-4659}
\affiliation{
  \institution{Universidade Federal de Pernambuco}
  \city{Recife}
  \country{Brazil}}
\email{anapaulaholanda.ap@gmail.com }

\author{Kiev Gama}
\orcid{https://orcid.org/0000-0003-1508-6196}
\affiliation{
 \institution{Universidade Federal de Pernambuco}
 \city{Recife}
 \state{Pernambuco}
 \country{Brazil}}
\email{kiev@cin.ufpe.br}
 
\author{Fernando Castor}
\orcid{https://orcid.org/0000-0002-6389-3630}
\affiliation{
  \institution{University of Twente}
  \city{Twente}
  \state{}
  \country{Netherlands}}
\email{f.castor@utwente.nl}
  
\author{Ronnie de Souza Santos}
\orcid{https://orcid.org/0000-0003-3235-6530}
\affiliation{
  \institution{University of Calgary}
  \city{Calgary}
  \state{Alberta}
  \country{CA}}
\email{ronnie.desouzasantos@ucalgary.ca}

\renewcommand{\shortauthors}{Santos et al.}

\begin{abstract}
\textbf{Context}: Diversity can impact team communication, productivity, cohesiveness, and creativity.  
%Diversity encompasses many dimensions, such as gender, ethnicity, and age. These aspects can cause individuals to be underrepresented or marginalized in teamwork.
Analyzing the existing knowledge about diversity in open source software (OSS) projects can provide directions for future research and raise awareness about barriers and biases against underrepresented groups in OSS. 
%It can also help us understand how these groups affect OSS projects. 
\textbf{Objective}: This study aims to analyze the knowledge about minority groups in OSS projects. We investigated which groups were studied in the OSS literature, the study methods used, their implications, and their recommendations to promote the inclusion of minority groups in OSS projects.
\textbf{Method}: To achieve this goal, we performed a systematic literature review study that analyzed \nbPapersFinalSelection papers that directly study underrepresented groups in OSS projects. 
\textbf{Results}:
Most papers focus on gender (62.3\%), while others like age or ethnicity are rarely studied. The neurodiversity dimension, have not been studied in the context of OSS.
Our results also reveal that diversity in OSS projects faces several barriers but brings significant benefits, such as promoting
safe and welcoming environments. 
\textbf{Conclusion}: Most analyzed papers adopt a myopic perspective that sees gender as strictly binary. Dimensions of diversity that affect how individuals interact and function in an OSS project, such as age, tenure, and ethnicity, have received very little attention. 
%Future research that covers dimensions beyond gender, such as disability and neurodiversity, can also be conducted to promote a more comprehensive understanding of diversity in the OSS ecosystem.
\end{abstract}


%\received{}
%\received[revised]{}
%\received[accepted]{}

\maketitle
\section{Introduction}

Diversity refers to the range of characteristics and cultural differences that distinguish individuals, making each person unique \cite{padilla2008social}. These differences contain various aspects of life and culture, and understanding them can be essential to progressing multiple sectors of society. Research has shown that diversity can positively impact teams and organizations \cite{stevens2008}, including software development teams \cite{catolino2019, vasilescu2015, rodriguez2021}.
For example, studies on diversity in OSS teams have identified a correlation between increased contributor diversity and enhanced team productivity \cite{vasilescu2015}.

However, the aspects characterizing the diversity of a team can also be aspects of social exclusion. For example, gender, sexual orientation, ethnicity, religion, disability status, location, employment status, age, and language ability could lead to social exclusion in the software engineering community (SE) \cite{10.1145/3593434.3593463}. Such social exclusion causes the emergence of groups that have historically faced marginalization and disadvantages due to characteristics such as gender, ethnicity, and sexual orientation. In this paper, we refer to these groups as underrepresented or ``minoritized'' groups \cite{wingrove2021not}, as they do not necessarily constitute numerical minorities.  

Recent research on diversity and inclusion in Software Engineering highlights challenges faced by underrepresented groups in OSS, such as negative experiences reported by women \cite{frluckaj2022} and bias in contribution evaluation based on ethnicity \cite{nadri2021}. However, prior studies often focus on specific aspects, such as the relationship between diversity and productivity \cite{vasilescu2015} or the prevalence of underrepresented groups in OSS \cite{zacchiroli2020}, leaving several diversity dimensions underexplored \cite{rodriguez2021}.  Therefore, we need a high-level perspective on the area to understand how various dimensions of diversity impact OSS development and how underrepresented groups participate in OSS projects. Hence, to gain a broader understanding of how various diversity dimensions impact OSS development and the participation of underrepresented groups, this paper presents a systematic literature review aiming to map existing studies, analyze research methods, identify gaps, and propose strategies to foster diversity and inclusion in OSS communities.

Our research builds on the work of Rodríguez-Pérez et al. \cite{rodriguez2021}, who conducted a systematic literature review to explore perceived diversity in SE. Their study identified key issues, methods, and tools while highlighting limitations in addressing diversity concerns in the field more broadly. Recognizing the challenges of OSS (e.g., maintainer bias, contributor onboarding difficulties, and community structures \cite{balali2018newcomers, guizani2021long}), our research extends and deepens this investigation by focusing specifically on diversity within OSS projects. 
Our work provides specific recommendations for fostering inclusion in open-source communities.

With this study, we provide the following key contributions: 1) An analysis of seven dimensions of diversity in the specific context of OSS (gender, sexual orientation, culture, ethnicity, age, disability, and neurodiversity); 2) An exploration of diversity trends in OSS, analyzing \nbPapersFinalSelection papers selected from \nbPapersFromAllEngines screened; 3) Evidence of the impact of diversity on OSS participation, showing that cultural and geographic diversity enhance participation and communication, racial and ethnic diversity lead to more frequent contributions, age diversity fosters engagement through intrinsic motivation, and cognitive diversity improves task performance; 4) The identification of structural challenges, where corporate restrictions and process complexity limit external participation, despite diversity promoting collaboration and retention.
\section{Background}\label{sec:background}

\subsection{Open source software (OSS)} 
\label{open-source-software}
OSS is software that can be executed, studied, modified, and distributed by its users for any goal\footnotemark{}. Availability of source code is a prerequisite for this. The openness of OSS is guaranteed by the use of OSS licenses.
This type of software has become a very important resource in information technology. For example, currently, the most widely used operating system in the world, Android, is distributed according to an OSS license. OSS is also very popular on the server side, in development tools, and in some other domains~\cite{fogel2005}.

OSS developers collaborate in a distributed, electronic, and asynchronous way \citep{eghbal2020}, allowing remote work from different parts of the world, using varied tools for communication and to coordinate work, without needing to be online simultaneously \cite{eghbal2020}. 
Successful open-source projects, such as Apache\footnotemark[\value{footnote}], often have formal governance structures where debates are resolved, new developers are invited, and new features are planned.

\footnotetext{Apache Software Foundation \url{https://www.apache.org/}}

Alternatively, they may use a less formal structure but more personal self-restraint on the part of leaders \cite{fogel2005}. These two structures primarily involve collaboration between developers. This collaboration can be affected by factors such as communication, biases, and discrimination \cite{vedres2019}.

Due to the distributed and open nature of OSS, projects have the potential to receive contributions from a diverse range of individuals. However, this diversity can reveal minoritized groups. For example, previous work \cite{ben2013} examined the contribution characteristics of developers in an open source environment based on visual analysis and revealed that some regions are more prevalent in an OSS project (e.g., USA, UK, and Germany) regarding contributions. While there is evidence suggesting gender diversity as a positive and significant indicator of productivity in OSS teams \cite{vasilescu2015}, the lack of gender diversity is a constant problem among OSS projects \cite{bosu2019}. 

\subsection{Diversity dimensions}
\label{diversity-dimensions}
 
Diversity is broadly defined as the presence of different characteristics in a group. In the workforce, it includes factors like age, culture, abilities, race, gender, and sexual orientation \cite{saxena2014}. Though originally from other fields, diversity has gained relevance in Software Engineering, where it enhances teamwork, creativity, and innovation \cite{roberge2010}.  

Diversity also has social implications, as some groups remain marginalized, reinforcing inequality \cite{neville2014racial}. These disparities extend to OSS communities. Our research adopts a broad view of diversity, using Saxena’s workforce diversity dimensions \cite{saxena2014}, widely recognized in team studies \cite{shore2009diversity}. These dimensions include: \textbf{Ethnicity:} Defined as shared ancestry, history, and cultural symbols contributing to identity \cite{cornell2006ethnicity}.  
\textbf{Culture:} Encompasses knowledge, beliefs, customs, and behaviors that shape social identity \cite{baecker1997}.  
\textbf{Gender:} Refers to roles and behaviors shaped by societal norms rather than biological sex \cite{peretti2015}.  
\textbf{Sexual Orientation:} Describes emotional, romantic, and sexual attraction to others \cite{stock2019}.  
\textbf{Age:} Relates to lifespan and changing interpretations of aging \cite{posner1995}.  
\textbf{Disability:} Involves physical or cognitive limitations affecting daily activities \cite{fitzpatrick1996}.  
\textbf{Neurodiversity:} Recognizes natural variations in brain function, including ASD, ADHD, and dyslexia \cite{sonuga2021neurodiversity}. Diversity strengthens teams by fostering different perspectives and skills \cite{roberge2010}. Our study focuses on minoritized groups in OSS, emphasizing those who face systemic barriers, not just numerical minorities.

\section{Methodology}
In this study, we aimed to identify the presence and participation of minoritized groups in OSS projects. We focused on studies that investigate the following dimensions: gender, sexual orientation, race, culture, age, disability, and neurodiversity. To achieve our goal, this study focused on answering four research questions:
\newcommand\rqone{Which underrepresented groups are studied in the context of OSS projects?}
\newcommand\rqtwo{Which methodological approaches are used to study underrepresented groups in OSS projects?}
\newcommand\rqthree{What is the experience of underrepresented groups in OSS projects?}
\newcommand\rqfour{What are the main practices or recommended solutions to include, engage, and onboard underrepresented groups in OSS projects?}
\begin{description}
    \item[RQ1] \rqone
    \item[RQ2] \rqtwo
    \item[RQ3] \rqthree
    \item[RQ4] \rqfour
\end{description}

Our research questions were designed to explore several aspects related to diversity in OSS. RQ1 investigates which dimensions of diversity are studied in papers on OSS. RQ2 question investigates the methodologies that current studies use, considering the types of studies and the methodologies they employ to identify members of underrepresented groups. RQ3 seeks to present the main results of existing studies about benefits and barriers faced by underrepresented groups. Finally, RQ4 aims to present the recommendations identified in studies on including minority groups in OSS projects. Therefore, this study can help us identify barriers that underrepresented groups face and obstacles to creating more diverse teams. 

\begin{figure}[t]
    \centering
    \includegraphics[width=0.426\textwidth]{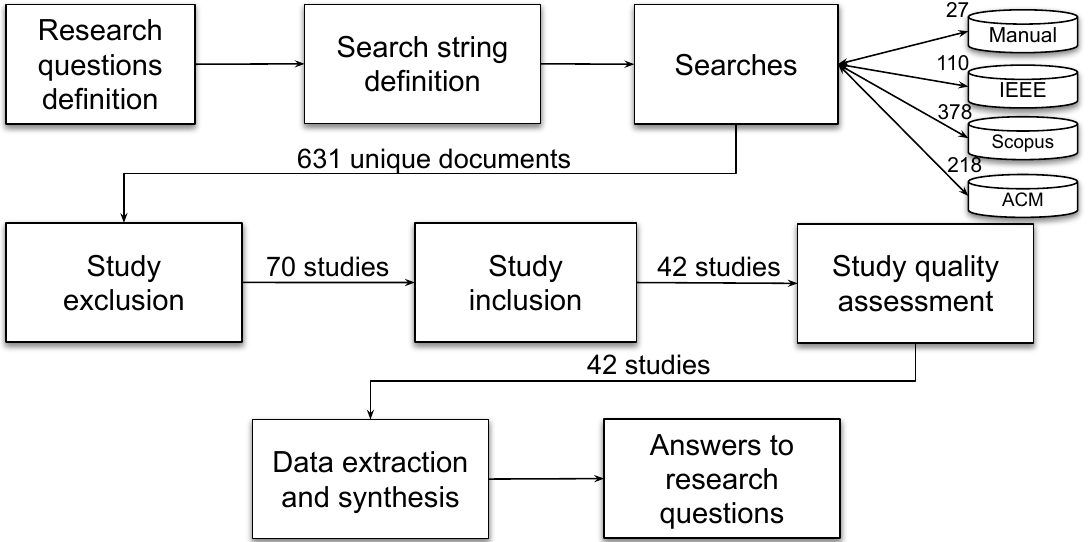}
    \caption{Systematic literature review roadmap.}
    \Description{Diagram illustrating the roadmap of a systematic literature review process.}
    \label{fig:mapping_roadmap}
\end{figure}

We followed the guidelines proposed by  \cite{kitchenham2015} and \cite{zhang2011identifying}. Figure \ref{fig:mapping_roadmap} presents the study roadmap. The methodology consisted of three phases. In the first phase, Planning, we defined the research question, developed the research protocol, which included search string, exclusion, and inclusion criteria, and validated the protocol. In the second phase, Conducting, we performed the study searching, applying the exclusion and inclusion criteria, and analyzing the resulting papers. The third phase corresponded to the documentation and dissemination of the results of this research. 
The next sections detail each step of the second phase.  

%https://docs.google.com/presentation/d/1cUb_dTGehhdyTsec9QUb35ZFrIF7TpLVWfb2XofFc28/edit?slide=id.g347878bbee3_1_79#slide=id.g347878bbee3_1_79

\subsection{Search Strategy}
Our strategy consisted of two parts: a manual search and an automated search. For the manual, the first author looked for papers in the top-tier software engineering conferences and jornals: ICSE, FSE, MSR, ASE, ESEM, ICSME, and SANER, CHASE, JSS, IST. \textcolor{black}{These venues were chosen because they are the main source of software engineering research, Additionally, they cover a wide range of topics. Journals selected for having high impact factors and a rigorous peer review process.} We limited this search to a specified time frame of the last 5 years. \textcolor{black}{The time limit was set to ensure the inclusion of recent advances.} In this search, the titles and abstracts were reviewed for relevance, and candidate papers were selected. The result of this process consisted of 27 papers.

For the automatic search, we built a search string using the terms of the dimensions of diversity investigated in this study and synonyms for underrepresented groups. We executed the search string in three search engines: the ACM (including the 
Guide to Computing Literature), IEEE, and Scopus. \textcolor{black}{Unlike manual search, this automated search was not limited to a time period. This allowed for the capture of a wider range of studies.} Our study utilized three digital libraries for two reasons. Firstly, the three digital libraries were selected based on their relevance to our research question and objectives. Secondly, using too many digital libraries can result in an excessive number of identical results, increasing the review process's workload and potentially making it infeasible.
The final search string used in these mechanisms and encompassing all planned diversity dimensions was:
\begin{center}
\footnotesize
Title~OR~Keywords with: ((``open source'') and (``divers*'' OR
``gender '' OR
``LGBT*'' OR
``queer'' OR
``sexual orientation'' OR
``race*'' OR
``ethnic*'' OR
``cultur*'' OR
``age'' OR
``disabilit*'' OR
``neurodivers*'' OR
``inclusi*'' OR
``equit*'' OR
``underrepresented groups'' OR
``marginalized groups'' OR
``social justice'' OR
``tenure*'' OR
``experienc*''))
\end{center}

This search string suffered formatting changes depending on the employed search engine. For example, for SCOPUS, we limited the search only to the computing area. However, none of the terms presented in the search string were discarded.  For some words of the search string, we did not use the '*' operator because they would return many false positives, that is, documents that do not fit within the scope of our study. For example, we performed the '*' wildcard in the word Age (ag*) in the ACM DL and produced 5,255 additional documents, many of which were about agility and age in contexts not related with software. This search process considered the years from 1990 to September 2024. We found \nbPapersFromAllEngines results.

\subsection{Study selection}

We performed an initial analysis of the \nbPapersFromAllEngines documents returned by the search to remove duplicates. This step analyzed only the document titles and keywords. After this filter, we had \nbPapersFromAllEnginesWithoutDuplicates ~documents. Then, the following exclusion criteria were applied:
\begin{itemize}
  \item EC1: Out of primary scope: The study is not related to underrepresented populations in OSS projects, and it’s not about software developers. 
  \item EC2: The study is not a full paper (e.g., master dissertations, doctoral theses, course completion monographs, short papers). As a rule of thumb, we consider that full papers must be at least five pages long. 
  \item EC3: The study cannot be downloaded directly from the libraries. This is ensured that the studies included in the research are readily available to other researchers who may want to review or build upon the findings.
  \item EC4: The paper did not undergo peer review, which is an essential part of the scientific and academic publishing process. It is a quality control mechanism in scientific research.
  \item EC5: Not written in English. English is the predominant language of scientific communication, and most scientific journals publish articles in English \footnote{Available in: \url{https://scientific-publishing.webshop.elsevier.com/manuscript-preparation/why-is-english-the-main-language-of-science/}}. 
  \item EC6: Duplicated and deprecated studies: Initially undetected duplicates and papers that were later superseded, e.g., because of an extended publication at a journal, are removed. In the latter case, only the most recent version is included. 
\end{itemize}

Short papers were omitted because they often lack the detail needed to conduct a more detailed analysis. \textcolor{black}{We focused on full-length articles to ensure that the evidence included was aligned with the study objectives. Short papers generally present preliminary results that are expanded in later papers.} All documents were analyzed by two authors. For this, the titles, abstracts and, in specific cases, the conclusions of the studies are analyzed. A paper is excluded from our working set when one or more exclusion criteria are satisfied. \textcolor{black}{As a reliability metric, we performed the kappa test, which had a result of 0.60, which can be considered a moderate agreement. In addition, all cases of disagreement were analyzed separately.} At the end of this stage, there were 70 papers.

After the exclusion step, we applied the inclusion criteria. Each paper must meet all the inclusion criteria to be considered acceptable for analysis. At this stage, the scope, objectives, and methodology of the 70 remaining studies from the exclusion stage were analyzed. The inclusion criteria used in this step were the following:
\begin{itemize}
    \item IC1: The study must investigate the practices, activities, individuals, tools, or artifacts connected to producing one or more OSS projects. 
    \item IC2: The study must investigate dimensions of diversity such as gender, LGBTQ, race, culture, age, disabilities, or neurodiversity. That is, the underrepresented population should be the object of study of the paper.
    \item IC3: Methodology: The study must be/contain at least one of the most relevant types of empirical studies in software engineering, such as controlled experiment, quasi-experiment, case study, survey research, interview study, ethnography, or action research.
    \item IC4: The paper focuses on activities that directly affect the source code and associated artifacts of an OSS project, e.g., feature implementation, bug fixing, testing, etc., and satellite activities where the main focus is the project, e.g., writing documentation (technical docs, scientific papers about the project), community outreach, and others.
\end{itemize}
IC1 is important because the open-source ecosystem involves a variety of stakeholders (developers, users, contributors, etc.), practices (coding, testing, reviewing, etc.), and artifacts (source code, documentation, issue tracker, etc.), and it is essential to study these to understand the dynamics of OSS projects.
IC2 defines the dimensions of diversity that this research investigates. By investigating the underrepresented groups, the study can shed light on the challenges and opportunities for making OSS projects more diverse and inclusive.
IC3 ensures that the study adopts a rigorous methodology that is appropriate for the research questions and objectives.
IC4 exists for two reasons: (i) OSS projects are primarily about software production, and any activity that affects the quality, usability, and sustainability of the software must be studied. (ii) OSS also involves building a community of contributors and users, and any activity affecting social and organizational aspects of the project should also be studied. Focusing on these activities can provide a broad understanding of the OSS project.

In this step, two researchers independently analyzed each study and categorized them as accepted, rejected, or possibly accepted. \textcolor{black}{We also calculated the kappa test which had a result of 0.061 meaning substantial agreement. Furthermore, all cases of disagreement were analyzed separately.} Applying the inclusion criteria and discussions lead to \nbPapersAfterInclusionCriteria ~papers for quality assessment. 

\subsection{Quality Assessment}

For the quality assessment, we used the instrument proposed by Dybå and Dingsøyr \cite{dybaa2008empirical}. This instrument’s intent is to assess three main issues: rigor, credibility, and relevance, and is composed of 11 questions. Each question should be answered with Yes/Partially/No. 
It is not the aim of this phase to exclude studies based on quality scores. Our aim is to assess the weight of the evidence found. These criteria provided a measure of the extent to which we could be confident in the contributions of the studies. The option ``Yes'' corresponds to 1, ``Partially'' corresponds to 0,5 and ``No'' to 0. The average of all values was $\approx 9.55$.

\subsection{Data Analysis}

We read the \nbPapersFinalSelection papers in full, extracted and analyzed the relevant data.  
For this phase, we employ a form with 14 questions to collect information about the studied subgroups, the type of study carried out, and the results found in the studies.

For the process of data synthesis, we initially performed a thematic analysis, guided by the Cruzes and
Dybå recommendations \cite{cruzes2011recommended}. This approach involves identifying, analyzing, and reporting patterns in textual data. The process began with familiarization with the data, followed by code generation. After this step, the codes were grouped into preliminary themes, which were refined to ensure coherence.
For this we analysis of the details of the publications, for example, the author's name, publication date, publication type, and publication source. So, for each research question, we analyzed different questions on the data extraction form. 
The protocol for conducting this study, as well as all the articles considered in each stage of this research, can be found at: \url{https://doi.org/10.5281/zenodo.7555875}.

\section{Results}

We found \nbPapersFinalSelection papers that studied minoritized groups in OSS projects. To better understand our data set, we organize our results using seven diversity dimensions discussed in \autoref{sec:background}. \autoref{tab:publications} presents the papers included in the study and the dimensions studied by each one of them. In this section, ``\textbf{Other}'' represents papers that address dimensions of diversity that we did not initially consider. For example, Daniel, Agarwal and Stewart~\citep{daniel2013} investigated the dimension of variety: role diversity across developers and users.

\begin{table*}[t]
\small
\begin{tabularx}{ \textwidth}{@{}p{0.5cm}Xl@{}}
\toprule
     
        \textbf{ID} & \textbf{Title} & \textbf{Year} \\
        \hline
        P01 & AID: An automated detector for gender-inclusivity bugs in OSS project pages & 2021 \\
        \hline
        P02 & An Empirical Investigation Into the Influence of Software Communities’ Cultural and Geographical Dispersion on Productivity & 2024 \\
        \hline
        P03 & Code reviews in open source projects: how do gender biases affect participation and outcomes? & 2023 \\
        \hline
        P04 & Codes of conduct in Open Source Software—for warm and fuzzy feelings or equality in community? & 2022 \\
        \hline
        P05 & Diversity and Inclusion in Open Source Software (OSS) Projects: Where Do We Stand? & 2019 \\
        \hline
        P06 & Floss (Free/Libre Open Source Software): A theme for studying cultural differences & 2008 \\
        \hline
        P07 & FLOSS participants' perceptions about gender and inclusiveness: a survey & 2019 \\
        \hline
        P08 & Gender and Participation in Open Source Software Development & 2022 \\
        \hline
        P09 & Gender and Tenure Diversity in GitHub Teams & 2015 \\
        \hline
        P10 & Gender differences and bias in open source: Pull request acceptance of women versus men & 2017 \\
        \hline
        P11 & Gender differences in public code contributions: A 50-year perspective & 2021 \\
        \hline
        P12 & Gender Representation Among Contributors to Open-Source Infrastructure: An Analysis of 20 Package Manager Ecosystems & 2023 \\
        \hline
        P13 & Gendered behavior as a disadvantage in open source software development & 2019 \\
        \hline
        P14 & Gendered Patterns of Politeness in Free/Libre Open Source Software Development & 2013 \\
        \hline
        P15 & Geographic diversity in public code contributions: an exploratory large-scale study over 50 years & 2022 \\
        \hline
        P16 & Going Farther Together: The Impact of Social Capital on Sustained Participation in Open Source & 2019 \\
        \hline
        P17 & How Gender-Biased Tools Shape Newcomer Experiences in OSS Projects & 2022 \\
        \hline
        P18 & Including Everyone, Everywhere: Understanding Opportunities and Challenges of Geographic Gender-Inclusion in OSS & 2022 \\
        \hline
        P19 & Investigating the effects of gender bias on GitHub & 2019 \\
        \hline
        P20 & Leveraging Corporate Engagement for Diversity in Free/Libre and Open Source Software Projects & 2023 \\
        \hline
        P21 & On older adults in free/open source software: reflections of contributors and community leaders & 2014 \\
        \hline
        P22 & On the Relationship Between the Developer’s Perceptible Race and Ethnicity and the Evaluation of Contributions in OSS & 2022 \\
        \hline
        P23 & Open source barriers to entry, revisited: a sociotechnical perspective & 2018 \\
        \hline
        P24 & Perceptions of the state of D\&I and D\&I initiative in the ASF & 2022 \\
        \hline
        P25 & Relationship between diversity of collaborative group members’ race and ethnicity and the frequency of their collaborative contributions in GitHub & 2023 \\
        \hline
        P26 & Social Diversity and Growth Levels of Open Source Software Projects on GitHub & 2016 \\
        \hline
        P27 & The effects of diversity in global, distributed collectives: A study of open source project success & 2013 \\
        \hline
        P28 & The State of Diversity and Inclusion in Apache: A Pulse Check & 2023 \\
        \hline
        P29 & Towards Understanding the Open Source Interest in Gender-Related GitHub Projects & 2023 \\
        \hline
        P30 & Understanding participation and corporatization in service of diversity in free/libre and open source software development projects & 2024 \\
        \hline
        P31 & Women participation in open source software communities & 2019 \\
        \hline
        P32 & Worldwide Gender Differences in Public Code Contributions and how they have been affected by the COVID-19 pandemic & 2022 \\
        \hline
        P33 & Designing for Cognitive Diversity: Improving the GitHub Experience for Newcomers & 2023 \\
        \hline
        P34 & From the Inside Out: Organizational Impact on Open-Source Communities and Women's Representation & 2024 \\
        \hline
        P35 & Discrimination, misogyny and harassment: Examples from OSS: content analysis of women-focused online discussion forums & 2022 \\
        \hline
        P36 & The State of Survival in OSS: The Impact of Diversity & 2023 \\
        \hline
        P37 & Work Practices and Perceptions from Women Core Developers in OSS Communities & 2022 \\
        \hline
        P38 & Gender Diversity and Women in Software Teams: How Do They Affect Community Smells? & 2019 \\
        \hline
        P39 & Onboarding vs. Diversity, Productivity and Quality — Empirical Study of the OpenStack Ecosystem & 2021 \\
        \hline
        P40 & Good fences make good neighbours?: on the impact of cultural and geographical dispersion on community smells & 2022 \\
        \hline
        P41 & Gender Differences in Early Free and Open Source Software Joining Process & 2012 \\
        \hline
        P42 & Women in Free/Libre/Open Source Software: The Situation in the 2010s & 2016 \\
\bottomrule
\end{tabularx}

\caption{Papers Included in this Systematic Review.}
\label{tab:publications}
\end{table*}

\subsection{RQ1 - Which underrepresented groups are studied in the context of OSS projects?}

The purpose of this research question is to investigate what are the dimensions of diversity found in studies of OSS projects. 
Our analysis reveals that 33 of the papers found in this study concentrated on gender diversity. Nine papers investigated other topics such as age, race, neurodiversity and other dimensions of diversity. No studies were found on disability and sexual orientation. 

\textbf{Age.} We found only one paper about this aspect. The paper investigate the barriers faced by older people in OSS communities (ID: P21). 
\textbf{Culture.} Four papers were found. Studies on culture focus on the impact of diversity on productivity (ID: P02), on the distribution and participation of developers (ID: P06), on perceptions of corporate involvement (ID: P20) and on communication and collaboration in distributed teams (ID: P40).

\textbf{Gender.} We found papers that address the challenges and barriers faced by women in OSS projects (IDs: P03, P08, P31 and P42). Others explore how OSS communities view diversity and inclusion initiatives (IDs: P05, P07, P12 and P35). Four studies investigate the impact of gender differences on participation and performance in OSS (IDs: P08, P09, P10 and P26). Two papers analyze how organizations can influence diversity and inclusion practices (IDs: P20 and P34). Finally, one study focuses on negative experiences and cases of discrimination (ID: P35). 

\textbf{Ethnicity.} We found 2 papers. One of these papers examines how the race and ethnicity of developers affect the evaluation of contributions to OSS projects (ID: P22). The other investigate the relationship between racial and ethnic diversity and the frequency of contributions to OSS (ID: P25).

\textbf{Other.} A set of seven articles investigated geographic dispersion and its relationship with productivity (IDs: P02, P26, P28 and P40) and gender inclusion (IDs: P18 and P30). We also found studies focused on cognitive diversity (ID: P33), disparity, separation, and variety (ID: P27), compensation (IDs: P28 and P36), social expectations and occupational biases (ID: P29), technical diversity, and corporate diversity (ID: P39).

\begin{summary}
Age, culture, gender, and race are essential dimensions studied in the literature. In addition, other aspects, such as cognitive style, geographic dispersion, and social expectations, have also been explored.  
\end{summary}

\subsection{RQ2 - Which methodological approaches are used to study underrepresented groups in OSS projects?}
Many research methods were found in the articles analyzed (e.g., surveys, case studies, mixed methods, quasi-experiments, diary studies, field studies). In total, 24 articles used a survey as the primary method. Surveys are often applied to large sample sizes from developers, with samples ranging from tens to thousands of participants. For example, P005 surveyed 4,543 developers from 10 popular projects. While P015 analyzed 2.2 billion commits from 43 million authors.

We found studies applied case study and interview methods. These studies usually include content analysis and data coding. For example, P35 analyzed 119 messages from 355 software packages in five forums, using NVivo for qualitative analysis. P008 conducted a case study with 23 interviews of female and male contributors in GitHub's PyPI ecosystem. Another seven studies adopted mixed approaches, combining quantitative (such as surveys) and qualitative (such as interviews or content analysis) in their analysis. For example, P16 analyzed quantitative data from 58,091 GitHub contributors and 88 qualitative survey responses. Meanwhile, P18 presented a mixed case study with quantitative data from the GHTorrent database and qualitative analysis.

Also we found studies used experimental or quasi-experimental methods to study the impact of different practices or behaviors on communities. These methods usually involve statistical modeling, multivariate regression, and data analysis. For example, P03 conducted a quasi-experiment using tools such as OpenCV and multivariate regression to study the influence of gender on code reviews. P38 conducted a quasi-experiment with 20 projects with all-male teams and 20 with mixed teams. 

Other studies used automated tools to classify developers' gender and ethnicity, such as GenderComputer, used in studies such as P09 and P10 to infer participants' gender. Namsor API and Name-Prism tool, used in such as P22 and P25, to infer ethnicity and gender based on contributors' names. \textcolor{black}{The analysis used in paper P22 is based on the U.S. Census Bureau’s use of six racial and ethnic groups: White, Black, Hispanic, API (Asian, Pacific Islander), AIAN (American Indian and Alaska Native), and 2PRACE (Mixed Race). While these categories provide a starting point, they are limited and do not fully capture the complexity of diversity in global OSS communities. These categories were created for a specific context in the United States and may not work well when applied to people from different regions.}
Other studies extract data from platforms (e.g., GitHub, StackOverflow), using APIs and scraping to collect large volumes of contribution data. P10 uses custom scraping to collect data on pull request status and user roles.

\begin{summary}
Most studies use surveys to collect quantitative data, with samples ranging from tens to thousands of participants. Some studies also employ automated tools to infer participants' gender and ethnicity based on public data. Mixed approaches, combining quantitative and qualitative data, are common. Experimental and quasi-experimental methods are less common.
\end{summary}

\subsection{RQ3 - Which is the experience of underrepresented groups in OSS projects?}

\textbf{\textit{Barriers.}}
Underrepresented groups in OSS face diverse barriers, with gender diversity being the most studied, particularly regarding women and cis people. Major obstacles include discrimination, gender bias, difficulties in code acceptance, delays, and limited review invitations (IDs: P03, P05, P10). Misogyny, sexism, and hostility create unwelcoming environments (IDs: P04, P35). Women also face social stigmas, struggles with mentorship, and challenges in proving competence and integrating into communities (ID: P07). High female turnover results from structural and cultural barriers (IDs: P11, P12, P41). Additionally, toxic environments, harassment, condescending comments, and unsolicited communication further drive exclusion (ID: P35).

Cultural and geographic diversity also presents significant challenges. Geographic dispersion and cultural diversity create communication difficulties in distributed teams, affecting collaboration and creating barriers to cultural integration (ID: P02). Geographic diversity can increase the effects of “Lone Wolf” and “Organizational Silo,” further complicating collaboration in multicultural teams (ID: P40). Furthermore, regions with higher 'Power distance' tend to have lower participation in OSS projects, and contributors from non-Western countries often disengage earlier due to lack of support and resources (IDs: P28, P36).

Regarding racial and ethnic diversity, developers perceptible as non-white face lower acceptance rates and are underrepresented in positions of influence in OSS communities (ID: P22). Racially and ethnically homogeneous groups, composed primarily of white developers, dominate the OSS space, which limits inclusion and diversity (ID: P25). Age diversity was also addressed, with one study focusing on contributors aged 50 and older (ID: P21). These individuals face social and technical challenges, such as conflicting expectations, difficulty understanding the codebase, and challenges with new tools (ID: P21).

Cognitive style is also a significant barrier. Participants with alternative cognitive styles need help with issues related to inclusion bugs, such as lack of feedback and confusing action options in interfaces, which disproportionately impact these individuals (ID: P33). Finally, structural and organizational challenges, such as the presence of corporate structures in OSS projects, can be seen as restrictive, limiting the flexibility of contributors (IDs: P20, P30). Demanding high-quality standards and a lack of transparency in contribution processes also hinder the engagement of external contributors (ID: P34).

\textbf{\textit{Benefits.}} 
Benefits of diversity dimensions are also evident. For gender, female contributions have a higher acceptance rate than male ones (ID: P10). Female representation is increasing in ecosystems like CRAN, showing greater receptiveness to gender diversity (ID: P12). Projects like Dreamwidth create inclusive environments that lower entry barriers and support women's participation (ID: P14). Community-building fosters ongoing engagement, with dedicated spaces providing emotional and technical support.
Cultural and geographic diversity is also growing, with more contributions from regions outside North America and Europe (ID: P15). This shift has promoted structured communication to overcome collaboration barriers like the ``Black Cloud effect'' (ID: P40).

Regarding racial and ethnic diversity, studies indicate that racially and ethnically diverse groups tend to contribute more frequently than homogeneous groups, suggesting a positive impact of racial diversity on the frequency of contributions (ID: P25). Results related to age diversity reveal that older contributors value the OSS community as a way to stay active and engaged, with motivations centered on pleasure and altruism (ID: P21). Interfaces designed to accommodate various cognitive styles have improved task completion rates, benefiting both participants and GitHub's inclusive environment (ID: P33).

Other benefits of diverse environments include skill development and networking opportunities. Gender diversity, combined with an inclusive culture, promotes more collaborative teams, creating a safer and more empathetic environment (IDs: P37, P38). In addition, diversity contributes to more remarkable survival and retention in corporate teams. Corporate-backed projects offer career opportunities, networking, compensation, and a more structured contribution experience, which can improve contributor retention (IDs: P20, P36). Finally, diversity also favors improved communication and reduced conflict, as the presence of women in OSS teams has been associated with a decrease in conflict (ID: P38).

\begin{summary}
Women in OSS face discrimination, misogyny, and recognition challenges, causing high turnover. Cultural and geographic diversity encounter communication issues and limited non-Western participation. Non-white developers face lower acceptance and lack representation in leadership. Age diversity hinders older contributors, and cognitive diversity complicates inclusion. Despite these challenges, diversity fosters higher acceptance of female contributions, greater global participation, frequent contributions from diverse groups, inclusive collaboration, skill growth, retention, and improved communication.
\end{summary}

\subsection{RQ4 - What are the main practices or recommended solutions to include, engage, and onboard underrepresented groups in OSS projects?}

The analysis results show that, among the \nbPapersFinalSelection papers selected for this study, only 6 provide explicit suggestions of practices, recommendations, or solutions. Within this subset, we found six papers (IDs: P14, P37, P31, P04, P18, P35) providing recommendations linked to the gender dimension, while only one (ID: P021) paper addresses the age dimension. We have come across broader recommendations as well which emphasizes the importance of promoting diversity (ID: P31). We classified the recommendations into seven groups: Better Environment, Self-Improvement, Recognition and Leadership, Evaluation Toolkit, Codes of Conduct, Support Network and Safe Spaces, and Mentorship.

Nine recommendations aim to create a \textbf{Better Environment} in a project to improve collaboration and foster diversity and inclusion. Some of this group's recommendations are abstract, e.g., creating or maintaining a healthy environment for underrepresented populations by adopting corrective action (ID: P04) and encouraging diversity. Others are more concrete and actionable, for example, setting quotas for women to increase their representation and understanding of the experiences of sexism and discrimination faced by them in OSS communities.

We also identified recommendations for \textbf{Self-Improvement}. This group includes recommendations related to personal empowerment. For example, ``develop a thick skin". This recommendation suggests that women should not be shaken by the criticism or negativity that may arise during their work on the project (ID: P31). It also includes recommendations for project members who do not belong to minoritized groups to improve, e.g., be supportive, friendly, respectful, and encouraging (ID: P31).

The seven recommendations about \textbf{Recognition and Leadership} deal with OSS participants' leadership skills and technical performance. For example, create initiatives to help develop women’s leadership skills and technical performance (ID: P37) and emphasize the success of female contributors both in women-focused spaces and in regular community communications to provide role models and encourage women's participation (ID: P04).

We placed four recommendations in the \textbf{Evaluation Toolkit} group. In this group are the recommendations to improve quality and inclusion in OSS projects. For example, (ID: P18) recommends providing a set of social metrics to help aspiring contributors gauge the quality of the community they are considering joining. (ID: P37), recommends continuous monitoring of female participation to generate metrics and track progress.

We found 11 recommendations for \textbf{Codes of Conduct}. Codes of conduct are widely used tools in OSS projects \cite{ tourani2017code}. The suggestions in this group include adopting a code of conduct in open software projects to ensure respectful and inclusive environments (P37) as well as the creation of codes of conduct (CoCs) that explicitly mention gender and/or minority status (IDs: P035, P031), (ID: P18). While most practices are related to creating CoCs, a few highlight the importance of enforcing them, e.g., enforcement of CoCs to ensure fairness and equality in a community (ID: P35).

Nine recommendations comprise the \textbf{Support Network and Safe Spaces}. This group gathers proposals aimed at making projects more welcoming for individuals of minoritized groups by reducing their sense of isolation and enabling them to exchange ideas comfortably. An example is the recommendation to build cohorts of older contributors through a private forum that will be a haven for older contributors to share their experiences (ID: P21).

Finally, the \textbf{Mentorship group} comprises four recommendations.  This group includes recommendations related to mentoring and providing guidance to project members. Two of these recommendations emphasize aspects that may be barriers to effective participation: culture and gender. For example, (ID: P18) recommends implementing proximity-based mentoring, where mentors and mentees are relatively close in region or culture. (ID: P31) recommends becoming mentors for women.

\begin{summary}
Improving diversity and inclusion in OSS needs effort and different strategies. Only six articles gave clear suggestions, mostly about gender, with one about age. These suggestions were grouped into seven categories:
Better Environment, Self-improvement, Recognition and Leadership, Assessment Kit, Codes of Conduct, Support Network and Safe Spaces, Mentoring.
\end{summary}

\section{Discussion}
In this section, we discuss our main findings. We begin by comparing our study with previous systematic literature reviews on diversity in software engineering. Next, we highlight the implications of our findings for both research and practice. Finally, we address potential threats to validity.

\subsection{Comparing Findings}
Our study presents similarities and reinforces previous findings in the literature, particularly regarding the emphasis on gender as the most studied diversity dimension in software engineering. Similar to Menezes and Prikladnicki~\cite{menezes2018}, Rodriguez et al.\cite{rodriguez2021}, and Prana et al.\cite{prana2021}, we found that gender diversity remains a central topic, with a significant focus on challenges faced by women in OSS. Our findings also align with prior research in recognizing barriers related to cultural and geographic diversity, as discussed by Rodriguez et al.\cite{rodriguez2021}. Additionally, our study confirms the methodological tendencies highlighted by Cosentino et al.\cite{cosentino2017}, as we also observed a strong reliance on surveys and mixed methods, with fewer experimental and quasi-experimental approaches being employed. These similarities suggest that the broader challenges of diversity in software engineering extend into OSS communities, reinforcing the need for continued focus on inclusion efforts.

On the other hand, our study diverges from previous systematic literature reviews in its scope and specific focus on OSS. While Menezes and Prikladnicki~\cite{menezes2018} examine diversity in software teams more broadly, our work is specifically concerned with how diversity is represented in OSS literature and how open-source contributors experience inclusion challenges. Similarly, while Rodriguez et al.\cite{rodriguez2021} distinguish between industrial and community-based environments, their study does not analyze OSS-specific factors such as maintainer bias, contributor retention, and governance structures, which are central to our findings. Moreover, prior research such as Zacchiroli et al.\cite{zacchiroli2020} and Prana et al.~\cite{prana2021} has largely focused on gender representation trends in OSS, whereas we take a broader perspective by examining multiple diversity dimensions, including race, age, and cognitive diversity.

Finally, we can highlight key contributions that make our study unique. Unlike previous reviews, we systematically map diversity research in OSS and identify significant gaps, particularly the lack of studies on disability and neurodiversity. While existing literature discusses diversity challenges in OSS, our study brings attention to the limited number of concrete strategies for fostering inclusion beyond gender diversity. By classifying existing recommendations and identifying underexplored areas, our work provides a foundation for future research aimed at developing targeted interventions that address a wider range of diversity dimensions in OSS communities.

\subsection{Implications to Research}
OSS projects face significant diversity challenges, and research must continue addressing gaps in understanding and improving inclusion. Previous studies have shown that OSS communities are predominantly male \cite{bosu2019, zacchiroli2020}. However, our findings highlight additional barriers, such as unintuitive interfaces and unclear documentation \cite{padala2020}, which may contribute to exclusion. These issues not only fail to support cognitive diversity but may also reinforce gender bias. Future research should explore how infrastructure and tool design affect participation, particularly for minoritized groups.

A major limitation in existing gender-focused studies is the reliance on a binary male/female framework. While some research has examined inclusivity for transgender individuals in hackathons and remote work \cite{prado2020, Ford:2019:HRW}, the roles and challenges faced by non-binary individuals in OSS remain largely unexplored. Expanding diversity research beyond binary gender perspectives would provide a broader understanding of inclusion in OSS communities.

Another critical issue is biased sampling. Previous studies emphasize the dominance of white men in OSS \cite{buffardi2016,nadri2021}, but this perspective overlooks significant global participation. Countries like China, India, Brazil, and Nigeria have large software development communities but are often underrepresented in diversity studies. Race is also a contextual concept; individuals categorized as white in Brazil, for example, may be considered Latino in the US \cite{Yancey:2003:WWL}. More research is needed to examine how race and ethnicity affect OSS participation across different cultural contexts.

The identification of minoritized groups in OSS also presents methodological challenges. Platforms like GitHub do not collect demographic data, making it difficult to assess diversity trends. Existing studies often rely on surveys, manual identification, or automated approaches, each with limitations. Automated tools frequently misclassify individuals, as shown in research on misgendering issues on Twitter \cite{FOSCHVILLARONGA:2021:LBT}. Privacy concerns also prevent many from disclosing sensitive information. Given regulations like GDPR, which prohibit processing demographic data without consent \cite{gdpr}, future research should explore ethical and privacy-conscious methods for assessing diversity in OSS.

Finally, while diversity-related challenges in OSS are well documented, most proposed solutions lack empirical evaluation. Many studies suggest recommendations, tools, and interventions, but few assess their effectiveness. Future research should systematically evaluate diversity strategies, such as mentorship programs, bias mitigation measures, and contribution recognition frameworks. Moreover, diversity studies in OSS rarely address disability and neurodiversity. Expanding research to include these dimensions is essential for fostering a more inclusive OSS ecosystem.

\subsection{Implications to Practice}
Promoting diversity in OSS is crucial for improving productivity, developer experience, and communication \cite{vasilescu2015, davidson2014, catolino2019}. Our findings suggest that targeted interventions can foster a more inclusive OSS environment. Encouraging broader participation, implementing and enforcing codes of conduct, and improving project infrastructure can lower barriers for underrepresented groups. Since unclear documentation and unintuitive interfaces contribute to exclusion \cite{padala2020}, making these elements more accessible can enhance cognitive diversity and reduce gender-related biases.

Codes of conduct play a fundamental role in setting behavioral expectations and promoting inclusive community values. Many OSS projects, including Linux and Apache HTTP Server, emphasize respect and inclusion in their codes of conduct\footnote{Linux code of conduct: \href{https://docs.kernel.org/process/code-of-conduct.html}{docs.kernel.org/process/code-of-conduct.html}} \footnote{Apache code of conduct: \href{https://www.apache.org/foundation/policies/conduct}{www.apache.org/foundation/policies/conduct}}. However, the effectiveness of these policies depends on consistent enforcement and adaptation to emerging challenges. Regularly updating codes of conduct based on community needs, as suggested by \cite{tourani2017code}, can help ensure their continued relevance.

Beyond codes of conduct, mentorship and diversity training support a more inclusive OSS culture. Safe spaces, recognition of diverse contributions, and structured mentorship encourage participation from women and other minoritized groups. Addressing bias through corrective action helps reduce contribution disparities. Maintainers should foster diverse participation and recognize diversity’s benefits, such as innovation and collaboration, to prioritize inclusion in team formation and reviews.

One of the most pressing gaps in OSS diversity efforts is the inclusion of minoritized groups beyond gender and race. While discussions often focus on increasing women’s participation, there is little practical guidance on supporting individuals with disabilities or neurodivergent. Incorporating accessibility measures into OSS development, such as designing inclusive tools and improving assistive technologies, can help bridge this gap. Similarly, ensuring that communication and collaboration mechanisms are adaptable to diverse needs can make OSS environments more welcoming.

As OSS projects grow in scale and influence, fostering diversity and inclusion is essential for long-term community health and innovation. Expanding efforts beyond gender and race, enforcing codes of conduct, and investing in accessible tools and mentorship can support more equitable OSS communities.

\subsection{Threats to Validity} 
As with any systematic literature review, this study has certain limitations and threats to validity associated with the chosen methodology. We discuss these limitations and the strategies implemented to mitigate them below.

\textbf{Construct validity.} 
This validity threat refers to ``identifying correct operational measures for the concepts under study" \citep{zhou2016map}.
We constructed our study from papers selected through a systematic process. Only three digital libraries were used to search for primary studies. This threatens construct validity because considering other digital libraries, such as Springer and Google Scholar, could result in a different sample of works. To mitigate this threat, we use the Guide to Computing Literature in the ACM digital library; this resource retrieves works published by other publishers. Furthermore, our seed studies were returned by ACM and IEEE. Finally, we did not use Google Scholar because the search string within that engine returned over 100,000 results from many different domains. 

\textbf{Internal validity.} 
This validity threat refers to ``seeking to establish a causal relationship, whereby certain conditions are believed to lead to other conditions, as distinguished from spurious relationships" \citep{zhou2016map}.
\textcolor{black}{One threat to internal validity lies in assessing the quality of the works involved in this research. This phase was conducted by the first author. To mitigate this thread, we adopt an instrument used and recommended by other studies.} Another potential threat is comparisons made with different studies due to variations in the methods and materials used in each research. To mitigate the first threat, we followed the best recommendations of the literature and 
used an popular instrument for evaluate the quality of the papers. Furthermore, we take a rigorous approach to analyzing and interpreting results to mitigate these two threats. The entire study selection process was conducted by at least two researchers. Next, when making comparisons between studies for grouping, contextual differences were considered. Finally, meetings were held to align the conclusions.

\textbf{External validity.} 
This validity threat ``Define the domain to which a study's findings can be generalized" \cite{zhou2016map}. 
Our study has focused on underrepresented groups in open-source communities and projects. Results and conclusions may not apply to other communities and projects or in another context. In addition, we only included in this study papers written in English. This decision could also be considered an internal threat, as relevant studies in other languages may have been excluded from this study.

\textcolor{black}{\textbf{Conclusion validity.} 
This validity threat "Demonstrate that the operations of a study such as the data collection procedure can be repeated, with the same results" \cite{zhou2016map}. In our study, two points are identified regarding this type of threat. The first corresponds to the data collection process. To ensure that the extracted data would be useful to achieve the research objective, we used a defined instrument that was validated by three authors of the study. The second corresponds to the data extraction process. For this stage, we also built an instrument and validated it with three collaborators. This instrument also allowed the responses to be verified.}

\section{Conclusions}
This work aimed to identify and classify existing studies investigating underrepresented groups in OSS. For this, we conducted a systematic literature review that analyzed \nbPapersFromAllEngines documents. Our study identified which underrepresented groups were investigated in previous papers, what were types of research carried out in this context, and what were the main conclusions of these studies. Most of the papers included investigated the gender aspect. Few papers were found that investigated dimensions of age and ethnicity. Papers about disability and neurodiversity were not found.

The main conclusions of the included studies show that older people seek to participate in OSS communities for reasons of identification with the community, altruism, and intrinsic motivation. The amount of contributions developed by Hispanic developers and Asian Pacific Islander (API) developers is low compared to white people. All studies focused on a binary male/female perspective of gender. Differences are found between the assessment of contributions of white individuals and individuals perceived as non-white. Finally, disparity diversity and tenure diversity are considered positive aspects of OSS projects. 

Our findings reveal gaps related to the dimensions of diversity investigated in the studies, the underrepresentation of ethnic groups in OSS projects, the limited variety of research methods used, understanding the impact of diversity, and exploring gender diversity beyond the binary model. Our study suggests that by implementing recommended strategies, such as creating an equitable environment and recognizing contributions, OSS communities can foster an inclusive culture that values the participation of minority groups.

\section*{Declaration of generative AI}
During the preparation of this paper the authors used ChatGPT to check spelling. After using this tool/service, the authors reviewed and edited the content as needed and take full responsibility for the content of the publication.

\bibliographystyle{ACM-Reference-Format}
\bibliography{bibliography}
\end{document}